\documentclass[a4paper,superscriptaddress,showkeys,prX,showpacs,twocolumn]{revtex4-1}
\usepackage{amssymb, amsmath, amsfonts,bm} %math formation
\usepackage{physics} %physics notation and symbols
\usepackage{siunitx} %add physics units
\usepackage{IEEEtrantools} %multiple lines equation
\usepackage{graphics,xcolor} %for graphics
\usepackage{epsfig}
\usepackage{pst-all}
\usepackage{hyperref} %add hyperlink to references
\usepackage[mathlines,running,modulo]{lineno}
\usepackage{relsize}
\usepackage[utf8]{inputenc}
\usepackage[toc]{appendix}

\newcommand{\varg}{\textsl{g}}
\newcommand{\ud}{\,\mathrm{d}}

\usepackage{dcolumn}
\usepackage{latexsym} %extra sybomls
\usepackage[version=3]{mhchem}

\begin{document}

%\preprint{APS/123-QED}

\title{A thorough investigation of the Antiferromagnetic Resonance}

\author{A. R. Moura}
\email{antoniormoura@ufv.br}
\affiliation{Departamento de Física, Universidade Federal de Viçosa, 36570-900, Viçosa, Minas Gerais, Brazil}

\date{\today}

\begin{abstract}
Antiferromagnetic (AF) compounds possess distinct characteristics that render them promising candidates for advancing 
the application of spin degree of freedom in computational devices. For instance, AF materials exhibit minimal susceptibility 
to external magnetic fields while operating at frequencies significantly higher than their ferromagnetic counterparts. 
However, despite their potential, there remains a dearth of understanding, particularly concerning certain aspects of AF spintronics.
In particular, the properties of coherent states in AF materials have received insufficient investigation, with many 
features extrapolated directly from the ferromagnetic scenario. Addressing this gap, this study offers a comprehensive 
examination of AF coherent states, shedding new light on both AF and Spin-Flop phases. Employing the Holstein-Primakoff 
formalism, we conduct an in-depth analysis of resonating-driven coherent phases. Subsequently, we apply this formalism 
to characterize antiferromagnetic resonance, a pivotal phenomenon in spin-pumping experiments, and extract crucial 
insights therefrom.

\end{abstract}


\maketitle

\section{Introduction and motivation}
\label{sec.introduction}
In recent years, there has been notable advancement in Condensed Matter Physics, 
particularly within the domain of spintronics research, indicating a forthcoming technological paradigm shift. 
Leveraging the capabilities of generating and controlling spin currents, there is anticipation for the emergence 
of a novel class of spintronic-based devices poised to supplant conventional electronic counterparts. Moreover, 
magnetic materials assume a central role in this pioneering endeavor. Notably, magnetic insulators hold particular 
intrigue for sustaining spin currents, as they mitigate energy losses and enable higher frequency operations 
compared to traditional electronic devices \cite{pr885.1}. Within this framework, Spin-Transfer Torque (STT) 
and Spin Pumping (SP) processes are frequently employed for the generation, manipulation, and detection of spin 
currents proximal to interfaces involving magnetic compounds. In the STT process, a spin current is injected into 
the insulator owing to the accumulation of electronic spin near the interface\cite{jmmm159.L1,prb54.9353}. 
Conversely, SP entails the generation of spin current by employing an oscillating microwave field, which 
provides angular momentum transferring to the material in contact with the magnetic insulator\cite{prl88.117601}. 

The SP mechanism entails the utilization of magnetic fields to induce a coherent precession of 
magnetization. This process involves the alignment of the spin field by a static magnetic field, complemented by 
an oscillating field that supports coherent dynamics. When the oscillating magnetic field is transversal to the 
static field,  we classify the process as linear, while for oscillating fields parallel to the static field, we 
have a nonlinear (or parametric) magnon excitation. By adjusting the intensity of the static magnetic field, we 
reach the resonance when the frequency of long wavelength magnons coincides with the oscillating field frequency. 
Consequently, there is an amplification in the population of low-energy magnons, which arise as spin currents 
into materials interfacing with the magnet.

In the course of magnetic resonance experiments, the spin field manifests synchronous dynamics, usually described 
through Coherent States (CS) formalism. Notably, coherent states, initially employed to conceive a comprehensive quantum model 
of radiation fields \cite{pr131.2766} and FM coherent magnons \cite{pla29.47,pla29.616,prb4.201}, are characterized 
by minimal uncertainty and are regarded as exhibiting the more classical-quantum nature \cite{rmp62.867}.
To illustrate, consider a particle confined within a harmonic potential, which is represented by a coherent state. 
In this context, the system holds a minimum uncertainty, $\Delta x\Delta p=\hbar/2$, and the wave function delineates a 
dispersionless wave packet that harmonically oscillates around the position of potential minimum. Analogously, the resonant spin 
field demonstrates semi-classical behavior, with classical fields depicted by the phase angle $\varphi$ around the 
$z$-axis, and the associated conjugate momentum denoted as $S^z$. This approach based on canonically conjugated fields
(or operators) is the leading feature of the Self-Consistent Harmonic Approximation (SCHA) \cite{jmmm472.1,prb106.054313}.

Though originally delineated for Ferromagnetic (FM) materials, both SP and STT processes exhibit analogous efficacy in 
Antiferromagnetic (AFM) insulators\cite{prl113.057601}. Historically, spintronics predominantly revolved around FM, 
with comparatively limited attention devoted to AFM. However, AFM spintronics has recently garnered significant interest, 
demonstrating superiority over FM applications \cite{natphy14.220,am2024.1521}. AFM insulators, for example, evince 
insensitivity to external magnetic perturbations and engender negligible stray fields due to the absence of macroscopic 
magnetization. Moreover, owing to AFM coupling, AFM frequencies extend to the terahertz (THz) range, while FM frequencies are 
restricted to the gigahertz (GHz) scale. A comprehensive exploration of AFM spintronics is available in various 
references \cite{ptrsa369.3098,natnano.3.231,rmp90.015005,rezende}. 

Despite the acknowledged significance of AFM materials in spintronic research, the existing framework for describing
Antiferromagnetic Resonance (AFMR) is deficient. Presently, much of the understanding regarding AFMR derives from a 
direct extension of the Ferromagnetic Resonance (FMR) results, yielding outcomes that are only partially accurate \cite{jap126.151101}. 
Notably absent in the current depiction of AFMR is an assessment of the coherence level, a factor critical for the 
applicability of Coherent States (CS). Consequently, a meticulous examination reveals a departure from the prevailing 
understanding in the literature, indicating that coherence in the Spin-Flop (SF) phase is confined to only one of 
the two AF modes. Furthermore, the determination of magnetic susceptibilities also necessitates precise knowledge of 
coherence levels. Therefore, in this work, we employ the Holstein-Primakoff (HP) formalism to present a thorough investigation 
of the AFMR. Our results contribute to a better understanding of the coherence dynamics of AF models, which is 
essential to many spintronic experiments.

\section{Semiclassical approach}
\label{sec.review}

In this section we present a brief review of the AFMR by using a semiclassical description. 
To begin with, we adopt a simple AF model given by the 
following Hamiltonian
\begin{equation}
\label{eq.hamiltonian}
	H = 2J \sum_{\langle i,j \rangle} \bm{S}_i^\prime \cdot \bm{S}_j^\prime - D \sum_i \left( S_{i}^{z\prime} \right)^2 - \gamma\hbar\sum_i \bm{B}_i^\prime (t)\cdot\bm{S}_i^\prime,
\end{equation}
where $J>0$ denotes the Antiferromagnetic exchange coupling constant, with $\langle i,j\rangle$ standing for 
nearest-neighbor spin interactions. $D>0$ represents a single-ion anisotropic constant, while 
$\bm{B}^\prime(t) = B_x^\prime(t) \hat{\imath}^\prime + B_y^\prime(t) \hat{\jmath}^\prime + B_z^\prime \hat{k}\prime$ 
denotes the magnetic field, and $\gamma=\varg\mu_B/\hbar=2\pi\times28$GHz/T is the gyromagnetic ratio. Note
that, for the sake of consistency, we assume that the spin and magnetic moment align in the same direction, without 
altering the final results. Furthermore, we apply prime notation to delineate spin components within the laboratory 
reference frame, whereas the absence of prime notation designates the local reference frame, as described below. 
Additionally, to simplify notation, directional indices ($x$, $y$, and $z$) will be interchangeably expressed as 
either subscript or superscript indices, with no substantive difference between 
the two representations.

For the sake of convenience, we establish the longitudinal direction, denoted as the $z$-axis, along the anisotropy 
axis, while the transverse components reside within the $xy$-plane. The static magnetic field $B_z^\prime$, pivotal 
in determining the phase exhibited by the model, aligns parallel to the anisotropic axis, whereas $B_x^\prime(t)$ and 
$B_y^\prime(t)$ represent the transverse components of the magnetic field. Consequently, we initially adopted an 
easy-axis model; nonetheless, the inclusion of easy-plane anisotropy necessitates minor adjustments in subsequent 
developments. Given the absence of substantial magnetization, resulting in the elimination of the demagnetizing field, 
we express $\bm{B}^\prime=\mu_0 \bm{H}_\textrm{ext}^\prime$, where $\bm{H}_\textrm{ext}^\prime$ denotes the external 
H-field. Generally, in spintronic experiments, we employ monochromatic oscillating fields characterized by a constant 
frequency $\Omega$. However, such consideration is not necessary from the theoretical perspective. Furthermore, owing 
to the diminished intensity of the transverse component, wherein $B_x^\prime\approx B_y^\prime\ll B_z^\prime$, we can 
treat the oscillating field as a perturbation.

\begin{figure}[h]
	\label{fig.phases}
	\centering 
	\epsfig{file=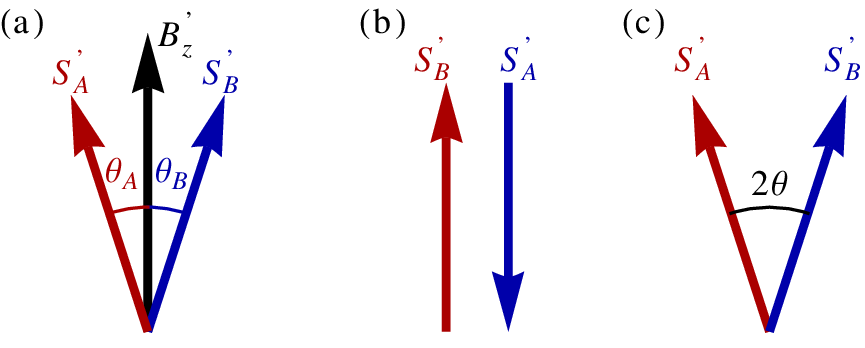,width=0.9\linewidth}
	\caption{(color online) (a) A general representation of the sublattice magnetizations. The red vector stands for the sublattice A, 
    and the blue represents the sublattice B. (b) For $B_z<B_\textrm{sf}$, we have the AF phase, which spin field solution is given 
    by $\theta_A=0$ and $\theta_B=\pi$. (c) For $B_z^\prime>B_\textrm{sf}$, the energy of the SF phase is minimized by the angles
    $\theta_A=-\theta_B=\cos^{-1}[B_z^\prime/(2B_E-B_D)]$.}
\end{figure}

Depending on the magnitude of $B_z^\prime$, two distinct phases emerge as a consequence of the energetic 
interplay between exchange and Zeeman energies. For a weak longitudinal magnetic field, the dominance of exchange 
energy results in the spin field assuming the conventional Antiferromagnetic (AF) phase, characterized by Néel 
ordering, where one sublattice magnetization opposes the direction of the other. Conversely, in the regime of strong 
magnetic fields, the Zeeman energy surpasses the exchange term, leading to the minimization of total energy in what is 
called the Spin-Flop (SF) phase. In the SF phase, sublattice magnetizations adopt the configuration depicted in 
Fig. (\ref{fig.phases}), with $\theta_A=-\theta_B$. 

To ascertain the critical field $B_\textrm{sf}$, which delineates the transition between the two 
phases, we consider the limit of uniform sublattice magnetization defined by $\bm{M}_l=\gamma\hbar\bm{S}_l/2V_c$, 
where  $V_c$ is the unit cell volume, and $l=A,B$ denotes the sublattice magnetization. Neglecting the 
influence of the small oscillating magnetic field, we derive an energy expression given by
\begin{IEEEeqnarray}{C}
	E_\textrm{cl}(\theta_A,\theta_B)=\frac{\gamma\hbar \mathcal{N}S}{2}\left[B_E \cos(\theta_A+\theta_B)-\frac{B_D}{2}(\cos^2\theta_A+\right.\nonumber\\
	\left.+\cos^2\theta_B)-B_z^\prime(\cos\theta_A+\cos\theta_B)\right],
\end{IEEEeqnarray}
where $\mathcal{N}$ denotes the number of spins and the exchange and the anisotropic fields are 
defined by $B_E=2zJS/\gamma\hbar$ and $B_D=2DS/\gamma\hbar$, respectively. Energy minimization yields the solutions 
$\theta_A=0$ and $\theta_B=\pi$ (representing the AF phase), or $\theta_A=-\theta_B=\arccos[B_z^\prime/(2B_E-B_D)]$ 
(characterizing the SF phase). Examination of the energy function reveals that the critical field is defined as 
$B_\textrm{sf}=\sqrt{2B_EB_D-B_D^2}$. Therefore, as the field intensity increases from small values, the N\'{e}el 
ordering decreases until the system undergoes a first-order phase transition at $B_z^\prime=B_\textrm{sf}$

Restricted to the constraint of a uniform spin field, corresponding to the $q=0$ magnons, the macroscopic magnetization 
demonstrates behavior analogous to that of a magnetic dipole undergoing precession under the influence of the static 
field $B_z^\prime$. Consequently, the magnetization dynamic is given by the Landau-Lifshitz (LL) equation, formulated as 
$d \bm{M}_l/dt=\gamma \bm{M}_l\times \bm{B}_{l,\textrm{eff}}$, where $\bm{B}_{l,\textrm{eff}}=-\partial H/\partial \bm{M}_l$ 
denotes the effective magnetic field acting upon sublattice $l$ (the negative sign is absence since we consider that 
spin and magnetic moment are parallel). By substituting the magnetization ansatz 
$\bm{M}_l(t)=M_l^z\hat{k}+(m_l^x\hat{\imath}+m_l^y\hat{\jmath})e^{-i\omega_0 t}$ into the LL equation, a system of coupled 
ordinary differential equations (ODEs) is obtained, yielding the $q=0$ magnon frequencies 
$\omega_0=\gamma B_z^\prime\pm \gamma (2 B_E B_D +B_D^2)^{1/2}$. In the absence of an external magnetic field, two 
degenerate modes emerge, precessing in opposing orientations. The frequency of the clockwise mode decreases with 
increasing $B_z^\prime$ and vanishes at the critical value $(2 B_E B_D +B_D^2)^{1/2}$, proximal to $B_\textrm{sf}$ for 
small anisotropies.

\section{Quantum description}
To properly describe the quantized spin-waves, we represent the spin operator by employing the HP representation.
As usual, to evaluate the magnon spectrum of Eq. (\ref{eq.hamiltonian}), it is useful to rotate the spins into a local 
reference frame, thereby establishing a collinear spin field. Under a generic scenario where each sublattice exhibits a 
distinct rotation angle, a rotation about the $x$-axis achieves the new spin fields in the local reference 
frame:
\begin{IEEEeqnarray}{rCl}
\label{eq.rotation}
	S_{r,l}^{x\prime} &=& S_{r,l}^x \nonumber\\
	S_{r,l}^{y\prime} &=& S_{r,l}^y \cos\theta_l - S_{r,l}^z \sin\theta_l \nonumber\\
	S_{r,l}^{z\prime} &=& S_{r,l}^z \cos\theta_l + S_{r,l}^y \sin\theta_l, 
\end{IEEEeqnarray}
where $r$ denotes the AF unit cell position and $l=A,B$ signifies the sublattice index. To avoid 
potential ambiguity, we reserve the indices $i,j$ for site position, which is equally represented by the index 
pair $r,l$. If we adopt $\mathcal{N}$ as the number of sites, then the number of unit cells is
given by $N=\mathcal{N}/2$. 

The development follows the usual procedures. Defining the quantization axis along the $z$-axis (in the local reference frame),  
we use the non-interacting HP representation to write the spin operators as $S_i^z=S-a_i^\dagger a_i$, $S_i^+=\sqrt{2S}a_i$, 
and $S_i^-=\sqrt{2S}a_i^\dagger$, for the sublattice A. Similarly, for the sublattice B, we achieve 
$S_j^z=S-b_j^\dagger b_j$, $S_j^+=\sqrt{2S}b_j$, and $S_j^-=\sqrt{2S}b_j^\dagger$. Here, $a$ and $b$ are bosonic operators
that satisfy $[a_i,a_j^\dagger]=[b_i,b_j^\dagger]=\delta_{i,j}$, and $[a_i,b_j^\dagger]=[a_i,b_j]=0$.
Adopting the low-temperature limit, we can disregard the magnon interactions, and expand the Hamiltonian to second order. 
Then, after a straightforward procedure, Eq. (\ref{eq.hamiltonian}) is expressed as $H(t)=E_0+H_1(t)+H_2$, where the constant term is given by
\begin{IEEEeqnarray}{rCl}
E_0&=&\gamma\hbar N\left[B_E(S+1)\cos\Delta\theta-\frac{B_D}{2}(S+1)(\cos^2\theta_A+\right.\nonumber\\
&&\left.+\cos^2\theta_B)-S(B_A^z+B_B^z)\right],
\end{IEEEeqnarray}
with $\Delta\theta=\theta_A-\theta_B$, and $B_l^z=B_z^\prime\cos\theta_l-B^{y\prime}\sin\theta_l$. 
Performing the Fourier transform $a_i=N^{-1/2}\sum_q a_q e^{i q r}$,
and an analogous equation for $b_i$, we obtain the time-dependent linear Hamiltonian in momentum space
\begin{equation}
H_1(t)=\sum_q[h_q^A(t) a_q^\dagger+h_q^B(t) b_q^\dagger]+\mathrm{h.c.},
\end{equation}
where the time-dependent coefficients are
\begin{IEEEeqnarray}{rCl}
\label{eq.hA}
h_q^A(t)&=&-\sqrt{\frac{S}{2}}\gamma\hbar[i(-B_E\sin\Delta\theta+\frac{B_D}{2}\sin2\theta_A)\sqrt{N}\delta_{q,0}+\nonumber\\
&&+B_{A,q}^+(t)],
\end{IEEEeqnarray}
and 
\begin{IEEEeqnarray}{rCl}
\label{eq.hB}
h_q^B(t)&=&-\sqrt{\frac{S}{2}}\gamma\hbar[i(B_E\sin\Delta\theta+\frac{B_D}{2}\sin2\theta_B)\sqrt{N}\delta_{q,0}+\nonumber\\
&&+B_{B,q}^+(t)].
\end{IEEEeqnarray}
In preceding coefficients, we define the magnetic field $B_{l,q}^+(t)=B_q^x(t)+i B_{l,q}^y(t)$, where the transformation
between the laboratory and local reference frames provides $B_q^x(t)=B_q^{x \prime}(t)$, and 
$B_{l,q}^y(t)=B_q^{y \prime}(t)\cos\theta_l+B_q^{z \prime}\sin\theta_l$ for $l=A,B$. As we will see, the linear 
Hamiltonian is directly responsible for the generation of coherent states, which sustain the magnetization precession. 
Finally, the quadratic Hamiltonian, which is time-independent, is expressed by
\begin{equation}
H_2=\frac{1}{2}\sum_q X_q^\dagger \mathcal{H}_q X_q,
\end{equation} 
where we define the vector $X_q^\dagger=(a_q^\dagger\ b_q^\dagger\ a_{-q}\ b_{-q})$, and the matrix
\begin{equation}
\label{eq.matrixH}
\mathcal{H}_q=\left(\begin{array}{cccc}
	A_a & B_q & C_a & D_q \\
	B_q & A_b & D_q & C_b \\
	C_a & D_q & A_a & B_q \\
	D_q & C_b & B_q & A_b 
	\end{array}\right)=I\otimes \mathcal{H}_q^{AB}+\sigma_x\otimes \mathcal{H}_q^{CD},
\end{equation}
whose coefficients are given by
\begin{IEEEeqnarray}{rCl}
	A_l&=&\gamma\hbar\left[-B_E\cos\Delta\theta+B_D\left(\cos^2\theta_l-\frac{\sin^2\theta_l}{2}\right)+\right.\nonumber\\
	&&\left.+B_z^\prime\cos\theta_l\right],  \nonumber\\
	B_q&=&\frac{\gamma\hbar}{2}(1+\cos\Delta\theta)B_E\gamma_q, \nonumber\\
	C_l&=&\frac{\gamma\hbar}{2}B_D\sin^2\theta_l,\nonumber\\
	D_q&=&\frac{\gamma\hbar}{2} (1-\cos\Delta\theta) B_E\gamma_q,
\end{IEEEeqnarray}
where $\gamma_q=z^{-1}\sum_{\delta r} e^{i\bm{q}\cdot\delta\bm{r}}$ is the structure factor of a lattice 
defined by $z$ nearest-neighbor spins located at $\delta\bm{r}$ positions.

In accordance with conventional procedures, the diagonalization of the Hamiltonian $H$ is achieved via a Bogoliubov transformation 
applied to the bosonic operators \cite{pr139.a450}. To this end, we introduce a new vector operator 
$\Psi_q^\dagger=(\alpha_q^\dagger\ \beta_q^\dagger\ \alpha_{-q}\ \beta_{-q})$, defined through the linear transformation 
$X_q=T_q\Psi_q$. In order to preserve the bosonic commutation relation, the matrix transformation must satisfy the 
condition $T_qGT_q^\dagger=G$, where $G=\sigma_z\otimes I$. Consequently, we have:
\begin{equation}
\label{eq.Hdiagonal}
X_q^\dagger\mathcal{H}_q X_q=\Psi_q^\dagger G(T_q^{-1}G\mathcal{H}_qT_q)\Psi_q=\Psi_q^\dagger\Omega_q\Psi_q,
\end{equation}
where $\Omega_q$ is a diagonal matrix given by $\Omega_q=\textrm{diag}(\omega_q^\alpha,\omega_q^\beta,\omega_q^\alpha,\omega_q^\beta)$.
Technically, the matrix $T_q$ diagonalizes $G\mathcal{H}_q$; however, determining the transformation for the general case, 
as defined by Eq. (\ref{eq.matrixH}), presents an intimidating challenge. Fortunately, under certain symmetrical conditions, 
exact solutions can be readily obtained for both the AF and SF phases.

Given that the transformation relations between $a_q$ and $b_q$, the more general transformation $T_q$ is expressed as
\begin{equation}
\label{eq.matrixT}
T_q=\left(\begin{array}{cc}
	U_q & V_q \\
	\bar{V}_q & \bar{U}_q
	\end{array}\right),
\end{equation}
where $U_q=[u_{mn}]$ and $V_q=[v_{mn}]$ are two-dimensional square matrices (the momentum index is implicit in coefficients).
Utilizing Eq. (\ref{eq.Hdiagonal}), it is determined that the submatrices of $T_q$ satisfy the following relations:
\begin{IEEEeqnarray}{rCl}
	\label{eq.UV}
	\IEEEyesnumber
	\IEEEyessubnumber*
	\mathcal{H}_q^{AB} U_q+\mathcal{H}_q^{CD}\bar{V}_q&=&U_q\Omega_q,\\
	\mathcal{H}_q^{AB} V_q+\mathcal{H}_q^{CD}\bar{U}_q&=&-V_q\Omega_q.
\end{IEEEeqnarray}
Expressing the new operators in terms of the coefficients $u_{mn}$ and $v_{mn}$ results in:
$a_q=u_{11}\alpha_q+u_{12}\beta_q+v_{11}\alpha_{-q}^\dagger+v_{12}\beta_{-q}^\dagger$, and
$b_q=u_{21}\alpha_q+u_{22}\beta_q+v_{21}\alpha_{-q}^\dagger+v_{22}\beta_{-q}^\dagger$. It is noteworthy that the 
matrix coefficients are not entirely independent, as the commutation relations between $a$ and $b$ impose constraints 
on them. For instance, the condition $[a_q,a_q^\dagger]=1$ implies $|u_{11}|^2+|u_{12}|^2-|v_{11}|^2-|v_{12}|^2=1$. 
Additionally, depending on the symmetries involved, it is feasible to parametrize the coefficients to simplify 
the diagonalization process.

\section{Magnon coherent states}
The static and dynamic constituents of the $H$ field, play a fundamental role in providing spin pumping 
driven by AFMR. In a standard AFMR experimental setup, an alternating magnetic field, oriented orthogonally to 
the static field, induces transversal oscillations of the spin field.
Throughout the experimental protocol, the frequency $\Omega$ of the oscillating field remains constant. Adjustments 
are then made to the static field to attain the resonance condition, which is achieved when the energy of 
the $q=0$ magnon energy equals $\hbar\Omega$. Upon successful fulfillment of the resonance 
condition, the entire spin field performs synchronous oscillations, thereby delineating a coherent magnetization state. 

The description of magnetization precession encounters limitations when exclusively framed in terms of energy 
(number) eigenstates. The transverse components of magnetization, depicted through first-order terms of the operators 
$a$ and $b$, yield $\langle S^x(t)\rangle=\langle S^y(t)\rangle=0$, where we employ the quadratic Hamiltonian $H_2$ for evaluating
the thermodynamic averages. To adequately capture resonant dynamics, the CS formalism
proves indispensable \cite{rmp62.867}. In essence, a coherent state $|\eta\rangle$ is defined as the eigenstate of the 
annihilation operator, denoted as $\alpha$, wherein $\alpha|\eta_\alpha\rangle=\eta_\alpha|\eta_\alpha\rangle$, and similarly 
for the $\beta$ operator. Alternatively, and equivalently, the coherent state can be expressed as 
$|\eta\rangle=D(\eta_\beta)D(\eta_\alpha)|0\rangle=D(\eta)|0\rangle$\footnote{The equivalence is valid since we are dealing 
with a model endowed with $SU(2)$ symmetry. For a more general model, only the definition based on the displacement operator 
shows correctness}, where $D(\eta_\alpha)=\exp(\eta_\alpha \alpha^\dagger-\bar{\eta}_\alpha \alpha)$ 
represents the $\alpha$-mode displacement operator, which provides $D^\dagger(\eta_\alpha)\alpha D(\eta_\alpha)=\alpha+\eta_\alpha$, 
and $|0\rangle$ denotes the vacuum state at $T=0$ K. The $\beta$-mode
displacement operator is defined likewise.

At finite temperatures, the thermodynamic properties of coherent states can be elucidated through the framework of Thermal 
Coherent States \cite{pla134.273}. An alternative and simpler approach to account for thermal effects within coherent states 
involves employing the density of states matrix denoted as $\rho_{CS}=D(\eta)\rho_0 D^\dagger(\eta)$, 
where $\rho_0=e^{-\beta H_2}/\textrm{Tr}(e^{-\beta H_2})$ \cite{jmo38.2339}. The statistical expectation value of an operator 
$O$ is subsequently computed as $\langle O\rangle=\textrm{Tr}(\rho_{CS}O)$. Utilizing the displacement operator, the expectation 
value $\langle \alpha^\dagger \alpha\rangle$ is derived as $|\eta_\alpha|^2+n_\textrm{th}(\epsilon)$, where $|\eta_\alpha|^2$ 
signifies the condensed component, and $n_\textrm{th}(\epsilon)$ represents the Bose-Einstein distribution of thermal states 
with energy $\epsilon$. This transition to the CS formalism offers an accurate depiction of the dynamics of 
coherent magnetization, thereby overcoming limitations associated with descriptions based solely on energy eigenstates.

The generation of coherent states is reached through the linear Hamiltonian $H_1$.
Given that $H_1$ is time-dependent, we employ the Interaction formalism to write the time evolution as 
$O(t)=S^\dagger(t)\hat{O}(t)S(t)$, where the caret denotes time evolution in accordance with the quadratic 
Hamiltonian $H_2$, and $S(t)=T_t\exp[-(i/\hbar)\int_0^t \hat{H}_1(t^\prime)dt^\prime]$ denotes the S-matrix. 
In this context, the time-ordering operator $T_t$ has a minor role and can be replaced by an irrelevant phase. 
Expanding $S(t,0)=T_t\prod_i S(t_i+\delta t,t_i)$, where $\delta t\ll 1$, and utilizing the 
Baker-Campbell-Hausdorff Formulae to multiply consecutive time-evolution operators, we derive:
\begin{equation}
\label{eq.S}
S(t)=e^{i\vartheta(t)}\exp\left[-\frac{i}{\hbar}\int_{-\infty}^\infty\hat{H}_1(t,t^\prime)dt^\prime\right],
\end{equation}
where $\vartheta(t)$ is a unimportant phase that cancels out after the averaging process. Additionally, we introduce 
the retarded Hamiltonian $\hat{H}_1(t,t^\prime)=\hat{H}_1(t^\prime)\theta(t-t^\prime)$ and implement an adiabatic time evolution 
to express $|\Psi\rangle=S(0,-\infty)|\Psi_0\rangle$, where $|\Psi\rangle$ and $|\Psi_0\rangle$ are the eigenstates
of the full and free Hamiltonians, respectively. It is noteworthy that, disregarding the irrelevant phase $\vartheta(t)$, the 
CS construction is exact and requires no weak interaction approximation.

Provided the diagonalized operator basis, it is a straightforward procedure to determine the coherent state eigenvalues 
$\eta_\alpha$ and $\eta_\beta$. In both AF and SF phase, we find that $S(t)=\prod_q D(\eta_q^\alpha(t))D(\eta_q^\beta(t))$, 
where $D(\eta_q^\alpha(t))=e^{\eta_q^\alpha(t)\alpha_q^\dagger-\bar{\eta}_q^\alpha(t)\alpha_q}$, and similarly for $D(\eta_q^\beta)$.
Then, applying the displacement operator, it is direct to obtain $\langle \alpha_q(t)\rangle=\langle D^\dagger(\eta_q^\alpha(t))\hat{\alpha}_q(t)D(\eta_q^\alpha(t))\rangle=\eta_q^\alpha(t)e^{-i\omega_q^\alpha t}$. 
Note that, since $H_1$ depends linearly on the bosonic operators, the CS formalism
yield a full description only of the linear magnon excitation, while the parametric scenario should be solved by a 
different approach. In the following sections, we utilize the CS framework to examine AFMR in both AF and SF phases, 
offering a more comprehensive description than that found in the current literature \cite{jap126.151101}, 
which revels previously unknown features.

\section{AF phase}
\textbf{Energy spectrum} - In the AF phase, the sublattice angles are given by $\theta_A=0$ and $\theta_B=\pi$, which vanish the coefficients $B_q$, $C_a$, and $C_b$
in the Hamiltonian matrix (\ref{eq.matrixH}). The new operators $\alpha_q$ and $\beta_q$ are then obtained from the matrix 
transformation $U_q=\cosh\Theta_q I$, and $V_q=-\sinh\Theta_q \sigma_x$, which results in
\begin{IEEEeqnarray}{rCl}
	\label{eq.bogoliubovAF}
	\IEEEyesnumber
	\IEEEyessubnumber*
	a_q&=&\cosh\Theta_q\alpha_q-\sinh\Theta_q\beta_{-q}^\dagger,\\
	b_q&=&\cosh\Theta_q\beta_q-\sinh\Theta_q\alpha_{-q}^\dagger.
\end{IEEEeqnarray}
The angle $\Theta_q$ (defined to be positive) is properly chosen to eliminate the Hamiltonian off-diagonal terms, which is reached when
\begin{equation}
	\label{eq.ThetaAF}
	\tanh2\Theta_q=\frac{2D_q}{A_a+A_b}=\frac{B_E \gamma_q}{B_E+B_D}.
\end{equation}
After replacing Eq. (\ref{eq.ThetaAF}) in the Hamiltonian, we achieve the diagonal model
$H_{AF}=E_0^{AF}+\sum_q(\epsilon_q^\alpha \alpha_q^\dagger\alpha_q+\epsilon_q^\beta \beta_q^\dagger \beta_q)$, where
$E_0^{AF}=E_0+\sum_q(\hbar\omega_q/2)$ is a constant energy, 
\begin{IEEEeqnarray}{rCl}
	\label{eq.H1H2}
	\IEEEyesnumber
	\IEEEyessubnumber*
	\epsilon_q^\alpha&=&\hbar\omega_q+\gamma\hbar B_z^{\prime},\\
	\epsilon_q^\beta&=&\hbar\omega_q-\gamma\hbar B_z^{\prime},
\end{IEEEeqnarray}
are the $\alpha$ and $\beta$-mode energies, while 
\begin{equation}
\omega_q=\gamma\sqrt{2B_E B_D+B_D^2+B_E^2(1-\gamma_q^2)}.
\end{equation}
As in the isotropic case, $E_0^\textrm{AF}<E_\textrm{cl}^\textrm{AF}$, and quantum fluctuations on the Neel ordering 
lead to a reduction of the ground state energy.

For small anisotropy field, the long-wavelength limit results in the relativistic dispersion relation
$\omega_q=(2/z)^{1/2}\gamma B_E aq$, where $a$ is the lattice spacing. Furthermore, it is straightforward procedure
to verify that the AF spin excitation exhibits integer spin aligns along the $z^\prime$ direction. 
This is evidenced by defining the (alpha) one-magnon eigenstate as follows 
\begin{equation}
\label{eq.psi}
|\alpha\rangle=\frac{1}{\sqrt{N}}\sum_q e^{i\bm{q}\cdot\bm{r}}\alpha_q^\dagger |0\rangle,
\end{equation}
where $|0\rangle$ is the vacuum state. Considering the total $S^z$ operator
\begin{equation}
S^{z\prime}=\sum_{i\in A} S_i^{z\prime}+\sum_{j\in B}S_j^{z\prime}=\sum_q (\beta_q^\dagger \beta_q-\alpha_q^\dagger \alpha_q),
\end{equation}
we found that $\langle\alpha|S^{z\prime}|\alpha\rangle=-1$ (in units of $\hbar$). A similar development reveals 
that $\langle\beta| S^{z\prime}|\beta\rangle=1$ for the beta mode $|\beta\rangle=N^{-1/2}\sum_q e^{iqr}\beta_q^\dagger|0\rangle$. 
Consequently, in the AF phase, the excitation of  microwave-driven magnons occurs through transverse oscillating fields, 
which couple the circularly polarized photons from the oscillating fields with the ladder spin operators. However, the inclusion of new terms in
Hamiltonian, such as the dipolar interaction, can alter this scenario and enable the excitation of 
magnons also by parallel oscillating fields \cite{prb21.7}.

\textbf{Coherent states} -  Examining the linear Hamiltonian reveals that the coefficients $h_q^A(t)$ and $h_q^B(t)$ 
assume the uncomplicated form 
$h_q^A(t,t^\prime)=-\sqrt{(S/2)}\gamma\hbar B_q^{+\prime}(t,t^\prime)$ and 
$h_q^B(t,t^\prime)=-\sqrt{(S/2)}\gamma\hbar B_q^{-\prime}(t,t^\prime)$, where
\begin{equation}
B_q^{\pm\prime}(t,t^\prime)=[B_q^{x\prime}(t^\prime)\mp iB_q^{y\prime}(t^\prime)]\theta(t-t^\prime),
\end{equation}
and we have included the Heaviside function to write the magnetic fields as retarded time functions. Performing
the Fourier transform in $t^\prime$, we obtain
\begin{IEEEeqnarray}{rCl}
\label{eq.B+}
B_q^{\pm\prime}(t,\omega)&=&\int_{-\infty}^t \ud t^\prime B_q^{\pm\prime}(t^\prime)e^{i(\omega-i\varepsilon)t^\prime}\nonumber\\
&=&\int\frac{\ud\nu}{2\pi}\frac{\tilde{B}_q^{\pm\prime}(\nu)}{i(\omega-\nu-i\varepsilon_q)}e^{i(\omega-\nu)t},
\end{IEEEeqnarray}
where we add the infinitesimal parameter $\varepsilon$ to ensure the convergence when $t^\prime\to-\infty$,
and $\tilde{B}(\nu)$ is the Fourier transform of $B(t^\prime)$. The convergence factor serves a similar purpose 
to that of a damping term, quantifying the dissipation of magnons under low-temperature conditions in 
the long-wavelength limit. In a more realistic model, the damping can be obtained from magnon interactions that
provides scattering processes to the spin-wave dynamics. However, for the present work purpose, it is sufficient to implement 
the damping by hand and consider experimental data for $\varepsilon$ when necessary. 

In terms of the $\alpha$ and $\beta$ operators, the argument of the exponential of $S(t)$ is given by
\begin{equation}
-\frac{i}{\hbar}\int_{-\infty}^\infty\ud t^\prime \hat{H}_1(t,t^\prime)=\sum_q (\eta_q^\alpha \alpha_q^\dagger+\eta_q^\beta \beta_q^\dagger)-\textrm{h.c},
\end{equation}
where 
\begin{equation}
\label{eq.etaalphaAF}
\eta_q^\alpha(t)=i\sqrt{\frac{S}{2}}e^{-\Theta_q}\gamma B_q^{+\prime}(t,\omega_q^\alpha)=|\eta_q^\alpha|e^{i\phi_q^\alpha},
\end{equation}
and
\begin{equation}
\label{eq.etabetaAF}
\eta_q^\beta(t)=i\sqrt{\frac{S}{2}}e^{-\Theta_q}\gamma B_q^{-\prime}(t,\omega_q^\beta)=|\eta_q^\beta|e^{i\phi_q^\beta}
\end{equation}
are the coherent state eigenvalues of the alpha and beta modes, respectively. In contrast to the scenario 
observed in FM systems, wherein solely one of the two circularly polarized photon modes interacts with the magnons, 
in the AF phase, it is feasible to induce microwave-driven magnons through the utilization of polarized magnetic fields 
exhibiting either left or right circular polarization. Additionally, linearly polarized fields can 
be employed to simultaneously excite both modes.

In the special case of a monochromatic circularly polarized field at frequency $\Omega$, the frequency coefficients 
are expressed as $\tilde{B}_q^{\pm\prime}(\nu)=2\pi B_q\delta(\nu\mp\Omega)$. In the low-temperature limit, thermal 
fluctuations are minimal and magnons predominantly arise from coherent behavior. Then, the monochromatic circularly 
polarized field provides straightforward outcomes for the populations of $\alpha$ and $\beta$ magnons. 
These populations are represented by
\begin{equation}
N_\alpha=\frac{S}{2}\sum_q e^{-2\Theta_q}\left|\frac{\gamma B_q}{\omega_q^\alpha-\Omega-i\varepsilon_q}\right|^2,
\end{equation}
and
\begin{equation}
N_\beta=\frac{S}{2}\sum_q e^{-2\Theta_q}\left|\frac{\gamma B_q}{\omega_q^\beta+\Omega-i\varepsilon_q}\right|^2,
\end{equation}
respectively. It's noteworthy that the momentum distribution is determined by the spatial behavior of the oscillating 
field. Typically, only one well-defined momentum $q_0$ significantly contributes to the magnon number. Furthermore, 
the condition $N_\alpha\gg N_\beta$ ($N_\alpha\ll N_\beta$) holds when $\Omega\approx\omega_q^\alpha$ 
($\Omega\approx-\omega_q^\beta$), resulting in the excitation of only one of the modes. For example, in the case of a 
uniform magnetic field $B_\textrm{rf}$ and clockwise orientation, the number of 
magnons can be expressed as
\begin{equation}
\label{eq.NmAF}
N_m^\textrm{AF}=\frac{e^{-2\Theta_0}}{2}\left(\frac{\gamma B_\textrm{rf}}{\varepsilon_0}\right)^2 N_\textrm{max},
\end{equation} 
where $N_\textrm{max}=NS$ is the maximum number of magnons in each sublattice. We obtain a similar outcome
for oscillating fields with clockwise orientation, and both modes are equivalent. Considering $e^{-2\Theta_0}=B_D/\sqrt{2B_EB_D}$, 
it becomes apparent that $N_m^\textrm{AF}$ exhibits no direct reliance on the static magnetic field. 
However, a negligible influence from $B_z^\prime$ may manifest in $\varepsilon_0$, given that the damping factor is contingent upon the 
magnon energy, which in turn is impacted by the static magnetic field. Consequently, an abrupt alteration in the coherence level is 
anticipated due to the first-order transition associated with the SF phase.

For the \ce{MnF_2}, an AF compound with $S=5/2$, the field parameters are $B_E=52.6$T and $B_D=0.82$T \cite{jmmm9.323}, 
while the damping factor $\varepsilon_0$ is of the order of $0.1$ GHz \cite{prb93.014425}.
Considering a stationary regime in which the absorption photon rate (proportional to $\gamma B_\textrm{rf}$) is
comparable in magnitude to the magnon damping, we get transverse magnetic fields at the scale of mT, typical
value for spintronic experiments. Adopting $\gamma B_\textrm{rf}\approx\varepsilon_0$, we obtain that $4.3\%$ 
of magnons are predicted in the coherent phase. It is worth noting that Eq. (\ref{eq.NmAF}) does not exhibit temperature dependence, 
which is atypical considering that thermal spin fluctuations typically diminish coherent behavior. The temperature 
dependence of magnon energies can be incorporated by utilizing renormalization parameters derived from higher-order 
interaction terms. For comparative purposes, we also obtained analogous outcomes by employing the SCHA formalism 
\cite{villela2024} to describe AFMR. The results derived from SCHA closely resemble those from the HP formalism, and 
in the low-temperature regime, the harmonic approximation yields $N_m^{AF}/N_\textrm{max}\approx 0.04$, a value in good 
accordance with the aforementioned result.

Parallel to the coherence level, we can determine the correlation function expressed by 
$C(\bm{r},\bm{r}^\prime)=\langle S^-(\bm{r})S^+(\bm{r}^\prime)\rangle=2S\langle a^\dagger(\bm{r})a(\bm{r}^\prime)\rangle$,
which describes the spatial relationship between spin operators.
In a normal state (non-coherent), it is expected that the correlation vanishes as $\Delta\bm{r}=\bm{r}-\bm{r}^\prime$ 
approaches to infinity. However, in the coherent state, we find a finite correlation despite the separation $\Delta\bm{r}$. Indeed, for the alpha mode,
we obtain
\begin{equation}
C(\Delta\bm{r})=\sum_q \frac{S^2\cosh^2\Theta_q}{Ne^{2\Theta_q}}\left|\frac{\gamma B_q}{\omega_q^\alpha-\Omega-i\varepsilon_q}\right|^2 e^{i\bm{q}\cdot\Delta\bm{r}}.
\end{equation}
For an almost uniform magnetic field, $B_q=\sqrt{N}B_\textrm{rf}\delta_{q,0}+\delta B_q$, where $\delta B_q$ represents 
the deviations from the uniform magnetic field.  At the resonant state, the correlation function is simplified, 
yielding both a baseline contribution and a deviation term, given by 
$C(\Delta\bm{r})=S^2(2B_D+B_\textrm{sf})/4B_\textrm{sf}+\delta C(\Delta\bm{r}),$
where we adopt $\gamma B_\textrm{rf}\approx\varepsilon_0$, and 
\begin{equation}
\delta C(\Delta\bm{r})=\sum_q \frac{S^2\cosh^2\Theta_q}{Ne^{2\Theta_q}}\left|\frac{\gamma \delta B_q}{\omega_q^\alpha-\Omega-i\varepsilon_q}\right|^2 e^{i\bm{q}\cdot\Delta\bm{r}}.
\end{equation}
The smoothness of $\delta B_q$ with respect to $q$ ensures that $\delta C(\Delta\bm{r})$ tends to zero for large
separations. Therefore, a finite correlation is observed even for large separations. 
Furthermore, the system exhibits Off-Diagonal Long-Range Order (ODLRO), a typical characteristic of condensed quantum states, 
due to the influence of the oscillating magnetic field. The same feature is observed in the SF phase, albeit with a 
different level of condensation.

\textbf{AF resonance} - By employing the CS description, we can determine the magnetization precession dynamics and 
consequently the proprieties of the AFMR. To begin with, we evaluate the average of $S_q^{+\prime}(t)=S_q^{x\prime}(t)+iS_q^{y\prime}(t)$.
For the sublattice A, the relevant averages depend on $a$ operator and, applying the Bogoliubov transformation, we achieve 
$\langle a_q(t)\rangle=\cosh\Theta_q \eta_q^\alpha e^{-i\omega_q^\alpha t}-\sinh\Theta_q \bar{\eta}_q^\beta e^{i\omega_q^\beta t}$, 
where the coherent state eigenvalues are time-dependents. Therefore, we obtain 
$\langle S_{A,q}^{+\prime}(t)\rangle=\sqrt{2S}[\cosh\Theta_q \eta_q^\alpha e^{-i\omega_q^\alpha t}-\sinh\Theta_q \bar{\eta}_q^\beta e^{i\omega_q^\beta t}]$. 
Since one of the coherent modes is established, the averages of the transverse spin component exhibit a 
simple sinusoidal behavior, similar those observed in semiclassical analysis. For instance, adopting the $\alpha$-mode, 
we found $\langle S_{A,q}^{x\prime}(t)\rangle=\sqrt{2S}\cosh\Theta_q|\eta_q^\alpha|\cos(\omega_q^\alpha-\phi_q^\alpha)$
and $\langle S_{A,q}^{y\prime}(t)\rangle=-\sqrt{2S}\cosh\Theta_q|\eta_q^\alpha|\sin(\omega_q^\alpha-\phi_q^\alpha)$. Moreover, 
in the AF phase, the spin circular dynamics shows no ellipticity, which results in a time-independent average of the 
$S_q^z$ component. Indeed, from the Bogoliubov transformation, it is straightforward to show that 
$\langle S_q^{z\prime}\rangle$ is constant in time when a coherent mode is defined. Note that, many general aspects of 
the magnetization precession can be obtained without the determination of the coherent state eigenvalues. These outcomes 
coincide to those obtained from the semiclassical approach, which takes into account only a uniform spin field. However,
for a more detailed investigation, the evaluation of $\eta_q^\alpha$ (and $\eta_q^\beta$) is essential, as it is explored below.

In particular, we are focused on employing the coherent states formalism to determine the magnetic susceptibility, 
as it plays a crucial role in resonance experiments, providing insights into magnetic damping through power absorption. 
Taking into account only the magnetic component of the microwave radiation, the absorbed power is expressed as
$P=(1/2)\textrm{Re}\int dV\bm{\bar{H}}\cdot \partial\bm{B}/\partial t.$
In momentum space, $\bm{B}_q(t)=\mu_0\bm{H}_q(t)+\mu_0\int\ud t^\prime \chi_q(t-t^\prime)\bm{H}_q(t^\prime)$, 
where $\chi_q$ denotes the (retarded) susceptibility tensor. For a linearly polarized field $H_q^y$, 
the power can be straightforwardly evaluated as
\begin{equation}
P=\frac{\mu_0\Omega}{2}\sum_q \textrm{Im}\chi_q^{yy}(H_q^y)^2,
\end{equation}
and, adopting a uniform field, $H_q^y=\sqrt{V}H^y\delta{q,0}$. The eigenvalues of coherent states offer a 
direct assessment of the magnetic susceptibility. Moreover, contrary to the LLG approach, which only provides 
insights into uniform magnetization precession, our findings allow for a comprehensive analysis across the entire 
Brillouin zone.

For determining the magnetic susceptibilities, we define the sublattice magnetization as 
$M_{A,q}^+(t)=(g\mu_B/a^3)\langle S_{A,q}^{+\prime}(t)\rangle$, which results in
\begin{IEEEeqnarray}{rCl}
M_{A,q}^+(t)&=&i\gamma M_s e^{-\Theta_q}\left[\cosh\Theta_q B_q^{+\prime}(t,\omega_q^\alpha)e^{-i\omega_q^\alpha t}-\right.\nonumber\\
&&\left.-\sinh\Theta_q \left(B_q^{-\prime}(t,\omega_q^\beta)e^{-i\omega_q^\beta t}\right)^\ast\right],
\end{IEEEeqnarray}
where $M_s=g\mu_B S/a^3$ is the saturation magnetization. It is convenient express the magnetization in frequency space and, 
after performing the Fourier transform in $t$, we obtain
\begin{equation}
\int\ud t B_q^{+\prime}(t,\omega_q^\alpha)e^{-i(\omega_q^\alpha-\Omega)t}=\frac{\tilde{B}_q^{+\prime}(\Omega)}{i(\omega_q^\alpha-\Omega-i\varepsilon_q)},
\end{equation}
which provides
\begin{IEEEeqnarray}{rCl}
\tilde{M}_{A,q}^+(\Omega)&=&\gamma M_s e^{-\Theta_q}\left(\frac{\cosh\Theta_q}{\omega_q^\alpha-\Omega-\varepsilon_q}-\right.\nonumber\\
&&\left.-\frac{\sinh\Theta_q}{\omega_q^\alpha+\Omega+\varepsilon_q}\right)\tilde{B}_q^{+\prime}(\Omega).
\end{IEEEeqnarray}
For the sublattice $B$, the procedures are similar and $\tilde{M}_{B,q}^+(\Omega)$ is obtained from $\tilde{M}_{A,q}^+(\Omega)$ replacing
$\cosh\Theta_q$ by $-\sinh\Theta_q$, and vice-versa. The total magnetization $M_q^+=M_{A,q}+M_{B,q}$ is then express as
\begin{equation}
\tilde{M}_q^+(\Omega)=\frac{2\gamma^2 M_s[B_D+B_E(1-\gamma_q)]}{(\omega_q^\alpha-i\varepsilon_q-\Omega)(\omega_q^\beta+i\varepsilon_q+\Omega)}\tilde{B}_q^{+\prime}(\Omega).
\end{equation}
For AF models, it is reasonable to assume small susceptibility and write $\bm{B}\approx\mu_0\bm{H}$. Therefore, for a circularly polarized
magnetic field, the susceptibility is given by
\begin{equation}
\tilde{\chi}_q^+(\Omega)=\frac{2\gamma\omega_s[B_D+B_E(1-\gamma_q)]}{(\omega_q^\alpha-i\varepsilon_q-\Omega)(\omega_q^\beta+i\varepsilon_q+\Omega)},
\end{equation}
where $\omega_s=\gamma \mu_0 M_s$. The susceptibility retrieve the well-known outcome of 
Keffer and Kittel \cite{pr85.329} for the $q=0$ mode. 

One can apply similar steps for determining the susceptibility associated with a linearly polarized field. However, provided the
symmetry between the transverse directions, it is simpler to use the relation $\tilde{M}_q^+=\tilde{M}_q^x+i\tilde{M}_q^y$, where  
$\tilde{M}_q^x=\tilde{\chi}_q^{xx} \tilde{H}_q^x+\tilde{\chi}_q^{xy} \tilde{H}_q^y$, and 
$\tilde{M}_q^y=\tilde{\chi}_q^{yy} \tilde{H}_q^y+\tilde{\chi}_q^{yx} \tilde{H}_q^y$. The symmetry ensures that
$\tilde{\chi}_q^{xx}=\tilde{\chi}_q^{yy}=(\tilde{\chi}_q^+ +\tilde{\chi}_q^-)/2$, and
$\tilde{\chi}_q^{yz}=-\tilde{\chi}_q^{xy}=(\tilde{\chi}_q^+ -\tilde{\chi}_q^-)/2i$.
Additionally, since $\chi^+(t)=[\chi^+(t)]^\ast$, we have $\tilde{\chi}_q^-(\Omega)=[\tilde{\chi}^+(\Omega)]^\ast=\tilde{\chi}^+(-\Omega)$, and so
\begin{IEEEeqnarray}{rCl}
\tilde{\chi}_q^{xx}(\Omega)&=&\frac{-2\gamma\omega_s[B_D+B_E(1-\gamma_q)]}{[(\omega_q^\alpha-i\varepsilon_q)^2-\Omega^2][(\omega_q^\beta+i\varepsilon_q)^2-\Omega^2]}[\Omega^2-\nonumber\\
&&-(\omega_q^\alpha-i\varepsilon_q)(\omega_q^\beta+i\varepsilon_q)],
\end{IEEEeqnarray}
while 
\begin{IEEEeqnarray}{rCl}
\tilde{\chi}_q^{yz}(\Omega)&=&\frac{-4i\gamma\omega_s[B_D+B_E(1-\gamma_q)]}{[(\omega_q^\alpha-i\varepsilon_q)^2-\Omega^2][(\omega_q^\beta+i\varepsilon_q)^2-\Omega^2]}(\gamma B_z^\prime-\nonumber\\
&&-i\varepsilon_q)\Omega,
\end{IEEEeqnarray}
For the uniform scenario, $q=0$, we again retrieve the established results of Keffer and Kittel.

\section{SF phase}
When dealing with the SF phase, $\theta_A=-\theta_B=\theta=\arccos[B_z^\prime/(2B_E-B_D)]$, for $\sqrt{2B_E B_D}<B_z^\prime<2B_E-B_D$. 
Therefore, the matrix coefficients are simplified to
$A_a=A_b=A=\gamma\hbar[-B_E\cos 2\theta+B_D(2\cos^2\theta-\sin^2\theta)/2+B_z^\prime\cos\theta]$, $B_q=\gamma\hbar B_E\gamma_q\cos^2\theta$, 
$C_a=C_b=C=\gamma\hbar (B_D\sin^2\theta)/2$, and $D_q=\gamma\hbar B_E\gamma_q \sin^2\theta$. By demanding positive 
angle in the long-wavelength limit, the Hamiltonian is diagonalized by the following transformation
\begin{equation}
\label{eq.matrixUsf}
U_q=\frac{1}{\sqrt{2}}\left(\begin{array}{cc}
	\cosh\Xi_q & \cosh\Phi_q \\
	-\cosh\Xi_q & \cosh\Phi_q
	\end{array}\right),
\end{equation}
and
\begin{equation}
\label{eq.matrixVsf}
V_q=\frac{1}{\sqrt{2}}\left(\begin{array}{cc}
	\sinh\Xi_q & -\sinh\Phi_q \\
	-\sinh\Xi_q & -\sinh\Phi_q
	\end{array}\right).
\end{equation}
Note that $a_q$ and $b_q$ involves the creation and annihilation operators from both alpha and beta
modes. Consequently, it is not simple to define the spin of the mode created by $a_q^\dagger$ (or $b_q^\dagger$). 
Employing the above parametrization, the angles $\Xi_q$ and $\Phi_q$ are obtained by solving 
Eq. (\ref{eq.UV}), which results in
\begin{IEEEeqnarray}{rCl}
	\label{eq.XiPhi}
	\IEEEyesnumber
	\IEEEyessubnumber*
	\tanh\Xi_q&=&\sqrt{\frac{A-B_q-\epsilon_q^\alpha}{A-B_q+\epsilon_q^\alpha}},\\
	\tanh\Phi_q&=&\sqrt{\frac{A+B_q-\epsilon_q^\beta}{A+B_q+\epsilon_q^\beta}},
\end{IEEEeqnarray}
while magnon energies are given by
\begin{IEEEeqnarray}{rCl}
	\label{eq.energySF}
	\IEEEyesnumber
	\IEEEyessubnumber*
	\epsilon_q^\alpha&=&\sqrt{(A-B_q)^2-(C-D_q)^2},\\
	\epsilon_q^\beta&=&\sqrt{(A+B_q)^2-(C+D_q)^2}.
\end{IEEEeqnarray}

\begin{figure}[h]
	\label{fig.angles}
	\centering 
	\epsfig{file=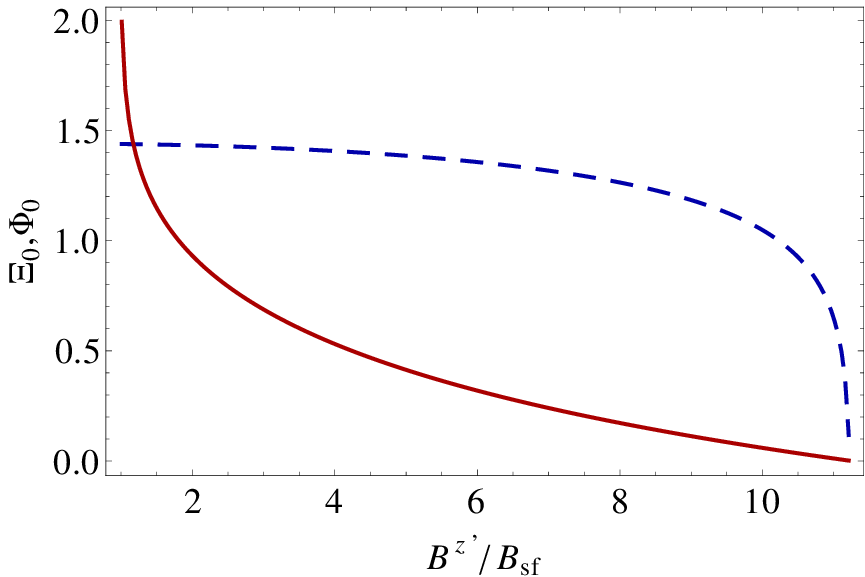,width=\linewidth}
	\caption{(color online) $\Xi_0$ (blue dashed line) and $\Phi_0$ (red solid line) are decreasing functions
	of increasing magnetic field. The angles are maximum at $q=0$ and tend to zero at the boundary of the first
	Brillouin zone. The angles were evaluated using the \ce{MnF_2} data.}
\end{figure}

Fig. (\ref{fig.angles}) depicts the diminishing trend of $\Xi_0$ and $\Phi_0$ as the static magnetic field increases. 
Both angles converge to zero as $B^{z\prime}$ approaches $2B_E-B_D$, implying that $a_q=(\alpha_q+\beta_q)/\sqrt{2}$ and 
$b_q=(-\alpha_q+\beta_q)/\sqrt{2}$. Consequently, in the limit of a strong magnetic field, only one type of spin fluctuation exists, 
and both modes $a_q$ and $b_q$ define FM magnons, as detailed below. 

In a result analogous to the previously observed in the AF phase, the diagonal Hamiltonian characterizing the SF phase is 
written as $H_{SF}=E_0^{SF}+\sum_q(\epsilon_q^\alpha \alpha_q^\dagger\alpha_q+\epsilon_q^\beta \beta_q^\dagger \beta_q)$, 
where $E_0^{SF}=E_0+\sum_q(\epsilon_q^\alpha+\epsilon_q^\beta)/4<E_\textrm{cl}^\textrm{SF}$. After the SF phase transition, evident 
distinctions emerge in the behavior of the alpha and beta modes. The alpha mode displays a gapless relativistic 
energy spectrum, in contrast to the beta mode which manifests a gapped quadratic dispersion relation akin 
to the fluctuations inherent to the ferromagnetic model (in the presence of an external magnetic field).
Indeed, in the long-wavelength the $\alpha$-mode energy is given by $\epsilon_q^\alpha=pc$, 
where $p=\hbar q$, and
\begin{equation}
c=\gamma a\sqrt{\frac{(B_D-2B_E)B_E\cos2\theta\sin^2\theta}{z}}
\end{equation}
is the spin-wave velocity, which simplifies to $c=\gamma B_E \sqrt{2/z}a$ when $B_D\ll B_E$. 
For the beta mode, we obtain $\epsilon_q^\beta=\epsilon_0+p^2/2m$, 
where the rest energy is given by $\epsilon_0=\gamma\hbar\sqrt{(2B_E\cos\theta)^2-2B_EB_D\sin^2\theta}$, and
\begin{equation}
m=\frac{z\sqrt{(2B_E\cos\theta)^2-2B_EB_D\sin^2\theta}}{2\gamma a[B_E^2(\sin^2\theta-2\cos^2\theta)+B_EB_D\sin^2\theta]}
\end{equation}
serves as a mass term (or gap energy). For small anisotropic fields, 
$m=(2z/\gamma a)\sqrt{(B_z^\prime)^2-B_\textrm{sf}^2}/[4B_E^2-3(B_z^\prime)^2]$.
Note that, at $B_z^\prime=B_\textrm{sf}$, the mass (and the rest energy) vanishes and the beta mode returns to the relativistic behavior, 
defining the AF phase. This distinct behavior between the alpha and beta mode energies directly impacts the 
coherence level within the system, leading to discernible differences in the magnetization precession behavior.
Furthermore, the gapless energy of the alpha mode results from the $O(2)$ symmetry of the spin field, which 
necessitates no energy cost to rotate the spin around the quantization axis. Introducing an easy-plane symmetry 
breaks this symmetry and opens a gap in the alpha mode as well.

\begin{figure}[h]
	\label{fig.spinSF}
	\centering 
	\epsfig{file=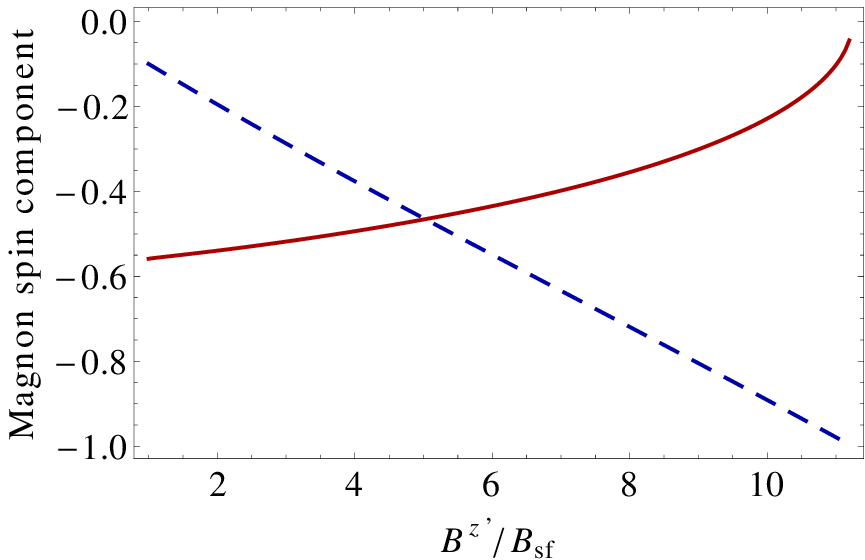,width=\linewidth}
	\caption{(color online) The magnon spin components in the SF phase vary as a function of the static magnetic field
	for \ce{MnF_2}. Here, the spin components are evaluated in the laboratory reference frame. 
	The solid red line corresponds to $\langle \delta S_A^{y\prime}\rangle_\alpha=-\langle \delta S_B^{y\prime}\rangle_\alpha$, 
	while the dashed blue line represents $\langle\delta S^{z\prime}\rangle_\alpha$. It is notable that 
	$\langle\delta S^{y\prime}\rangle_\alpha=\langle\delta S_A^{y\prime}\rangle_\alpha+\langle\delta S_B^{y\prime}\rangle_\alpha=0$. 
	Similar results are obtained for the $\beta$-mode.}
\end{figure}

Another notable distinction between the AF and SF phases arises in the 
spin orientations of the excitations. While the AF fluctuations exhibit spins aligned along 
the $z^\prime$ axis, for $B_z^\prime \gtrsim B_\textrm{sf}$, the SF magnons predominantly feature
spins oriented on the plane transverse to $z^\prime$. 
Defining $|\alpha\rangle$ similarly to the AF scenario and $S_A^z=\sum_{i\in A} S_i^z$ as the total spin $S^z$
 for sublattice A, we find that the magnon spin (in local reference frame) is given by
\begin{equation}
\langle \delta S_A^z\rangle_\alpha\equiv \langle\alpha| S_A^z|\alpha\rangle-\langle 0|S_A^z|0\rangle=-\sum_k \frac{\cosh 2\Xi_k}{2N}
\end{equation}
with an identical outcome for the sublattice $B$. 
In contrast to the AF scenario, in the SF phase, it is necessary to subtract the vacuum expected value $\langle 0|S_A^z|0\rangle$ to obtain a 
finite measurement of the spin fluctuation above the ground state background. The spin components $S^x$ and $S^y$ yield vanishing averages for 
both sublattices. Consequently, in the laboratory reference frame, the magnon spin exhibits components 
$\langle \delta S_A^{z\prime}\rangle_\alpha=\langle \delta S_B^{z\prime}\rangle_\alpha=\cos\theta \langle \delta S_A^z\rangle_\alpha$, and 
$\langle\delta S_A^{y\prime}\rangle_\alpha=-\langle \delta S_B^{y\prime}\rangle_\alpha=-\sin\theta \langle \delta S_A^z\rangle_\alpha$. 
Fig. (\ref{fig.spinSF}) illustrates the spin components 
$\langle \delta S^{z\prime}\rangle_\alpha=\langle \delta S_A^{z\prime}\rangle_\alpha+\langle \delta S_B^{z\prime}\rangle_\alpha$ and $\langle \delta S_A^{y\prime}\rangle_\alpha$ 
as a function of the static magnetic field $B_z^\prime$. For small values of $B_z^\prime$, the magnon spin components are approximately 
$\langle \delta S^{z\prime}\rangle_\alpha\approx 0.1$, while $\langle\delta S_A^{y\prime}\rangle_\alpha=-\langle\delta S_B^{y\prime}\rangle_\alpha\approx 0.55$ 
(in units of $\hbar$). Analogous outcomes for the $\beta$-mode are achieved by substituting $\Xi_q$ with $\Phi_q$. It is noteworthy that for 
magnetic fields near $2B_E$, the magnon adopts a state resembling the FM phase, in which both $\alpha$ and $\beta$-modes demonstrate the same 
magnon branch. Furthermore, at $B_z^\prime=2B_E-B_D$, 
the system undergoes a continuous phase transition to a paramagnetic state \cite{pr136.1068}. 
Generally, the second phase transition involves intense magnetic fields, and a detailed investigation of this phase falls outside the scope of 
the present work.

\textbf{Coherent States} - The scenario for the SF phase reveals numerous previously unidentified characteristics of 
magnon coherent states in the limit of strong magnetic fields. Diverging from what has been previously explored in literature, 
the SF coherent states do not constitute a direct extension of the AF case.  Indeed, as elucidated below, 
for strong magnetic field $B^{z\prime}$, the coherent state demonstrates behavior akin to the FM model.

In the SF phase, the coefficients of the linear Hamiltonian $H_1$ are given by 
$h_q^A(t,t^\prime)=\sqrt{S/2}\gamma\hbar[iB_0\delta_{q,0}-B_q^{x\prime}(t^\prime)+i\cos\theta B_q^{y\prime}(t^\prime)]\theta(t-t^\prime)$, and
$h_q^B(t,t^\prime)=\sqrt{S/2}\gamma\hbar[-iB_0\delta_{q,0}-B_q^{x\prime}(t^\prime)+i\cos\theta B_q^{y\prime}]\theta(t-t^\prime)$, where
$B_0=\sqrt{N}(-B_E\sin 2\theta+B_D\sin\theta\cos\theta+B_z^\prime\sin\theta)$ is a time-independent uniform term.

In terms of the $\alpha_q$ and $\beta_q$ operators, we obtain the linear Hamiltonian
\begin{IEEEeqnarray}{C}
\hat{H}_1(t,t^\prime)=-\sqrt{S}\gamma\hbar\theta(t-t^\prime)\sum_q\left\{ iB_0 e^{-\Theta_0}e^{i\omega_0^\alpha t^\prime}\delta_{q,0}\alpha_0^\dagger+\right.\nonumber\\
\left.+[e^{-\Phi_q} B_q^{x\prime}(t^\prime)+i\cos\theta e^{\Phi_q} B_q^{y\prime}(t^\prime)]e^{i\omega_q^\beta t^\prime}\beta_q^\dagger\right\},
\end{IEEEeqnarray}
which, after integration of the $t^\prime$ time, yields the coherent state eigenvalues
\begin{equation}
\label{eq.etaalphaSF}
\eta_q^\alpha=\frac{i\sqrt{S}\gamma B_0e^{-\Theta_0}}{\omega_0^\alpha-i\varepsilon_0}\delta_{q,0}e^{i\omega_0^\alpha t},
\end{equation}
and
\begin{equation}
\label{eq.etabetaSF}
\eta_q^\beta=i\sqrt{S}\gamma[e^{-\Phi_q} B_q^{x\prime}(t,\omega_q^\beta)+i\cos\theta e^{\Phi_q} B_q^{y\prime}(t,\omega_q^\beta)],
\end{equation}
where $B_q^{x\prime}(t,\omega_q^\beta)$ and $B_q^{y\prime}(t,\omega_q^\beta)$ are defined similarly to Eq. (\ref{eq.B+}). 
Opposite to the AF phase, for which the $\alpha$ and $\beta$ coherent states eigenvalues are similar, 
the coherence parameters $\eta_\alpha$ and $\eta_\beta$ exhibit distinct behavior. The oscillating magnetic field 
uniquely couples with the beta mode, whereas the alpha mode interacts solely with the static magnetic field.
In the absence of planar anisotropy, $\Theta_0$ tends towards infinity, and $\eta_\alpha$ approaches zero. With 
the inclusion of planar anisotropy, a gap arises in $\epsilon_\alpha$, and $\Theta_0$ becomes finite. However, 
the eigenvalue $\eta_\alpha$ is orders of magnitude smaller than $\eta_\beta$, rendering it negligible. Consequently,
the coherent state associated with the $\beta$-mode predominates.

To evaluate the coherence level, let us consider a monochromatic circularly polarized field $\bm{B}_\textrm{rf}^\prime$  with frequency $\Omega$. 
The Fourier transform of the transverse components are given by $\tilde{B}_q^{x\prime}(\nu)=\pi B_q[\delta(\nu-\Omega)+\delta(\nu+\Omega)]$,
and $\tilde{B}_q^{y\prime}(\nu)=i\pi B_q [\delta(\nu-\Omega)-\delta(\nu+\Omega)]$. Furthermore, when dealing with 
uniform magnetic fields, $B_q=\sqrt{N}B_\textrm{rf}\delta_{q,0}$, where $B_\textrm{rf}$ represents the magnetic field intensity. 
Therefore, in the resonant state of oscillating magnetic fields with clockwise orientation, which define 
photons with left circular polarization, the condition $\Omega=\omega_0^\beta$ provides
\begin{equation}
B_q^{x\prime}(t,\omega_q^\beta)=-iB_q^{y\prime}(t,\omega_q^\beta)\approx\frac{\sqrt{N}B_\textrm{rf}}{2\epsilon_0}\delta_{q,0}.
\end{equation}
From the semiclassical analysis, the clockwork resonating magnetic field induces a torque 
$\bm{\tau}_\textrm{rf}=\gamma\bm{m}\times\bm{B}_\textrm{rf}^\prime$ in the magnetization, opposing the damping torque. 
Furthermore, for $B^{z\prime}\gtrsim B_\textrm{sf}$, $\cos\theta e^{\Phi_0}\gg e^{-\Phi_0}$, we obtain a stronger magnetic 
field coupling along the $y$-direction than the $x$-direction, resulting in the elliptical magnetization precession. 
From the quantum perspective, left circular polarization photons, which carry angular momentum $-\hbar$, are absorbed by 
the magnetic sample, converting them into magnons. Therefore, the number of magnons stimulated by photons with 
left circular polarization contributing to coherence is determined as follows
\begin{equation}
N_m^\textrm{SF(L)}=|\eta_0^\beta|^2=\frac{1}{4}(e^{-\Phi_0}+\cos\theta e^{\Phi_0})^2 N_\textrm{max},
\end{equation}
where we adopt $\gamma B_\textrm{rf}\approx\varepsilon_0$. 

A parallel analysis demonstrates that photons exhibiting right circular polarization yield
\begin{equation}
N_m^\textrm{SF(R)}=\frac{1}{4}(e^{-\Phi_0}-\cos\theta e^{\Phi_0})^2 N_\textrm{max}.
\end{equation}
In this scenario, it is noteworthy that the absorbed photons carry spin $\hbar$, leading to an additional 
damping torque that increases with stronger static magnetic fields. When $B^{z\prime}\approx 2B_E-B_D$, the SF magnons possess 
spin $S^z=-\hbar$, resulting in the absence of photon-magnon interaction due to angular momentum conservation. 
Consequently, the stimulation of coherent magnons is significantly diminished. 
\begin{figure}[h]
	\label{fig.NmSF}
	\centering 
	\epsfig{file=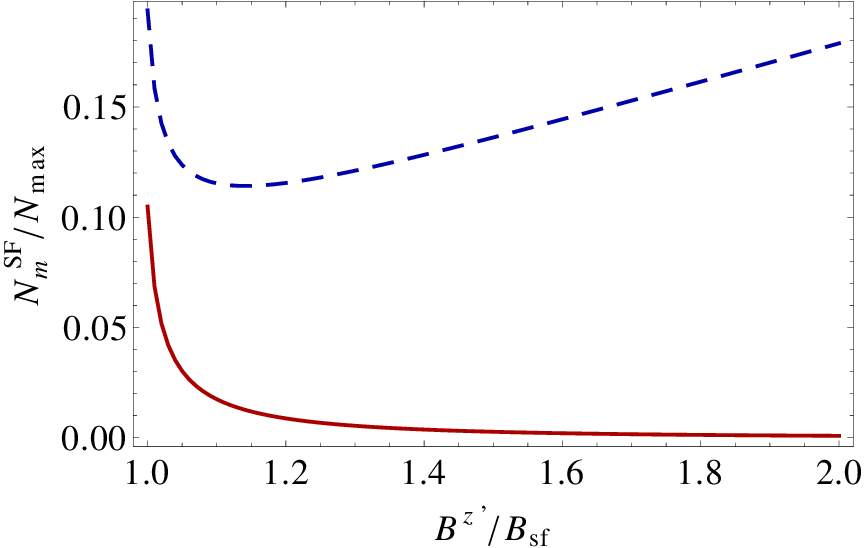,width=\linewidth}
	\caption{(color online) The number of magnon participating of the coherent state in the SF phase for
	\ce{MnF_2}. The (red) 	solid line represents $N_m^\textrm{SF(R)}$, while the (blue) dashed line 
	define $N_m^\textrm{SF(L)}$.}
\end{figure}

Figure (\ref{fig.NmSF}) illustrates the distinction between magnon stimulation by oscillating magnetic fields with 
left and right circular polarization for \ce{MnF_2}. The coherence level $N_m^\textrm{SF(L)}$ exhibits a minimum close to 
$B^{z\prime}=1.1B_\textrm{sf}$ owing to the differential coupling with the transverse components of the magnetic field. 
The coupling with $B_0^{y\prime}$ depends on $\cos\theta e^{\Phi_0}$, while $B_q^{x\prime}$ coupling involves the 
exponential term $e^{-\Phi_0}$. The exponential $e^{\Phi_0}$ exhibits rapid decline for 
$B_\textrm{sf}\leq B^{z\prime}\lesssim 1.1B_\textrm{sf}$, whereas $e^{-\Phi_0}$ is an increasing function. 
For $B^{z\prime}\gtrsim 1.1B_\textrm{sf}$, both couplings increase with the rising magnetic field, resulting in the increasing 
of the coherence level. When $B^{z\prime}\approx 2B_E$, the magnetization precession nearly becomes circular, 
and the entire spin field behaves akin to the FM model, with $N_m^\textrm{AF(L)}$ approaching $N_\textrm{max}$.

Utilizing the same dataset for \ce{MnF_2}, we determine $N_m^\textrm{SF(R)}=0.017N_\textrm{max}$ and 
$N_m^\textrm{SF(L)}=0.115N_\textrm{max}$ for $B_z^\prime=1.1 B_\textrm{sf}$ ($\approx 0.19 B_E$). 
In the low-temperature regime, SCHA formalism yields a similar outcome of $1.6\%$ of magnons in the 
coherent state for right circular polarization and $10.3\%$ for left circular polarization \cite{villela2024}. 

\textbf{SF resonance} - By employing the eigenvalues of coherent states, we can establish the dynamics of magnetization 
precession in the spin-flop (SF) phase, a crucial aspect in spintronic research. Furthermore, we derive the magnetic 
susceptibilities across the entire Brillouin zone. The methodology aligns closely with that employed in the AF phase.

To begin with, we utilize the outcomes from the preceding section to compute the averages of the spin components. 
Since $S^x$ and $S^y$ are solely comprised of linear combinations of HP bosonic operators, the averages can be readily 
evaluated. In the local reference frame, we obtain:
\begin{IEEEeqnarray}{rCl}
\langle S_q^x(t)\rangle&=&\langle\eta_q| [\hat{S}_{A,q}^x(t)+\hat{S}_{B,q}^x(t)]|\eta_q\rangle=\nonumber\\
&=&2\sqrt{S}e^{-\Phi_q}|\eta_q^\beta|\cos(\omega_q^\beta t-\phi_q^\beta),
\end{IEEEeqnarray}
and 
\begin{IEEEeqnarray}{rCl}
\langle S_q^y(t)\rangle&=&\langle\eta_q| [\hat{S}_{A,q}^y(t)+\hat{S}_{B,q}^y(t)]|\eta_q\rangle=\nonumber\\
&=&-2\sqrt{S}e^{\Phi_q}|\eta_q^\beta|\sin(\omega_q^\beta t-\phi_q^\beta).
\end{IEEEeqnarray}
As observed in the antiferromagnetic (AF) phase, the spin dynamics also exhibit precession around the $z$-axis; 
however, in the SF phase, the precession follows an elliptical trajectory with the semi-major axis aligned along the $y$-direction. 
Consequently, the total longitudinal spin component becomes time-dependent, expressed as:
\begin{IEEEeqnarray}{l}
S_q^z(t)=S-\frac{1}{2}\left[\cosh(2\Theta_q)\alpha_q^\dagger \alpha_q+\cosh(2\Phi_q)\beta_q^\dagger \beta_q+\right.\nonumber\\
+\frac{\sinh(2\Theta_q)}{2}\alpha_q^\dagger \alpha_{-q}^\dagger e^{2i\omega_q^\alpha t}+\frac{\sinh(2\Phi_q)}{2}\beta_q^\dagger \beta_{-q}^\dagger e^{2i\omega_q^\beta t}+\nonumber\\
\left.+\sinh^2\Theta_q+\sinh^2\Phi_q\right]+h.c.
\end{IEEEeqnarray}
Note that the off-diagonal term $\beta_q^\dagger \beta_{-q}^\dagger$ ensures parallel spin pumping when 
coupled with a time-dependent $B_q^z(t)$ magnetic field (alpha mode has negligible contribution). The efficiency of parallel spin 
pumping is contingent upon the angle $\Phi_q$ and diminishes in the limit of strong magnetic fields, where the precession 
returns to a circular orbit. Note that, when $B^{z\prime}\approx 2B_E$, both the local and laboratory reference frames tend to 
coincide, resulting in magnetization precession occurring along the direction of the static magnetic field.

To determine the transversal magnetic susceptibilities, we express the spin component averages in the
laboratory reference frame, which results in $\langle S_q^{x\prime}\rangle=\langle S_q^x\rangle$ and
$\langle S_q^{y\prime}\rangle=\cos\theta \langle S_q^y\rangle$, where we disregarded the $S^z$ terms, which
involve $\eta_\alpha$. Using Eq. (\ref{eq.etabetaSF}), we express the magnetization in frequency space as
\begin{equation}
\tilde{M}_q^x=2\gamma M_s\frac{\omega_q^\beta e^{-2\Phi_q}\tilde{B}_q^{x\prime}+i\Omega\cos\theta\tilde{B}_q^{y\prime}}{(\omega_q^\beta)^2-(\Omega+i\varepsilon_q)^2}
\end{equation}
where $M_s$ is the saturation magnetization. Following the same steps, we also obtain
\begin{equation}
\tilde{M}_q^y=2\gamma M_s\cos\theta\frac{\omega_q^\beta\cos\theta e^{2\Phi_q}\tilde{B}_q^{y\prime}-i\Omega\tilde{B}_q^{x\prime}}{(\omega_q^\beta)^2-(\Omega+i\varepsilon_q)^2}.
\end{equation}

In the context of the strong static magnetic field present in the SF phase, assuming the small susceptibility 
limit is not justified. Thus, we opt to express the magnetization equation using the relation $\bm{B}=\mu_0(\bm{M}+\bm{H})$, 
which leads to the matrix equation $\Gamma_q^M \tilde{M}_q=\Gamma_q^H\tilde{H}_q$. The susceptibility tensor 
$\chi_q$ is then determined as $(\Gamma_q^M)^{-1}\Gamma_q^H$, and its diagonal coefficients are straightforwardly 
evaluated as follows
\begin{equation}
\chi_q^{xx}=-\frac{\omega_s e^{-2\Phi_q}}{\Omega-\omega_q^\beta+\omega_s(e^{-2\Phi_q}+\cos^2\theta e^{2\Phi_q})+i\varepsilon_q},
\end{equation}
while $\chi_q^{yy}=(\cos\theta e^{2\Phi_q})^2\chi_q^{xx}$. Similarly, the off-diagonal susceptibilities are given by
\begin{equation}
\chi_q^{xy}=-\frac{i\omega_s \cos\theta}{\Omega-\omega_q^\beta+\omega_s(e^{-2\Phi_q}+\cos^2\theta e^{2\Phi_q})+i\varepsilon_q},
\end{equation}
and $\chi_q^{yx}=-\chi_q^{xy}$. In the above equations, we have employed the approximation 
$\omega_s\ll \omega_q^\beta$ to neglect terms involving $\omega_s^2$. Moreover, provided that the susceptibility is
apparent only in the vicinity of the resonant frequency, we have adopted the approximation
$\Omega^2-(\omega_q^\beta)^2\approx 2\Omega(\Omega-\omega_q^\beta)$ in the above development.
\begin{figure}[h]
	\label{fig.chi}
	\centering 
	\epsfig{file=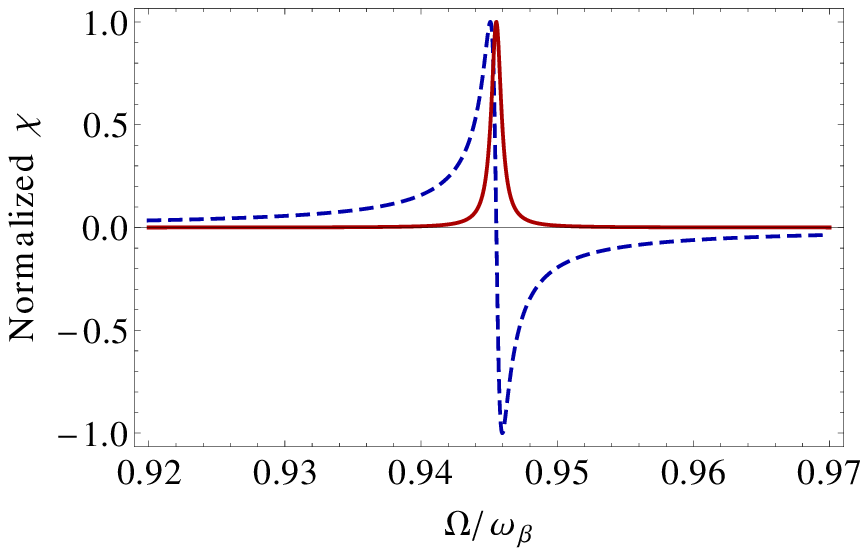,width=\linewidth}
	\caption{(color online) The real (blue dashed line) and imaginary (red solid line) part of the uniform 
	susceptibility ($\chi_{q=0}^{xx}$) for \ce{MnF_2} when $B^{z\prime}=1.1B_\textrm{sf}$.}
\end{figure}

It is noteworthy that the system exhibits greater responsiveness to oscillating magnetic fields applied along the 
$y^\prime$-direction, which aligns with the previous analysis of magnetization precession. Specifically, for 
$B^{z\prime}\gtrsim B_\textrm{sf}$, we observe $\chi_q^{yy}\approx 10^2 \chi_q^{xx}$, indicating a substantial difference 
between susceptibilities along different axes. However, as $B^{z\prime}$ approaches $2B_E-B_D$, this discrepancy diminishes, 
resulting in the absence of a preferential axis. Nevertheless, upon normalization, both $\chi^{xx}$ and $\chi^{yy}$
susceptibilities exhibit remarkably similar characteristics. Fig. (\ref{fig.chi}) shows the real and imaginary components 
of the normalized uniform susceptibility for \ce{MnF_2}. Notably, the SF susceptibilities mirror the findings observed in 
ferromagnetic resonance.

\section{Summary and Conclusion}

In this article, our focus lies on the examination of coherent states and resonance phenomena in antiferromagnetic models. 
While the equivalent problem in ferromagnetic systems has been extensively studied and documented, the AF model presents a
distinct scenario. Despite this difference, many characteristics of coherent states in AF systems have been determined by a 
direct extension of those observed in FM systems \cite{jap126.151101}.

We have conducted a detailed investigation of coherent states in AF models using the Holstein-Primakoff formalism. 
Although alternative spin representations exist, the HP approach is widely utilized due to its simplicity, particularly 
in describing ordered phases. Through a linear spin-wave approximation, we have provided a comprehensive account of the 
generation of coherent states induced by oscillating magnetic fields in both AF and SF phases. While the outcomes in the AF 
phase align with existing literature, our study has revealed novel findings in the SF phase, contrary to current understandings. 
Notably, we have observed that only one of the two AF modes exhibits significant coherent behavior, and the SF phase exhibits a 
closer resemblance to FM behavior than AF behavior. Leveraging the insights gained from coherent states, we have further 
elucidated the properties of AFMR, including magnetic susceptibilities.

In summary, our study contributes to bridging the gap in the understanding of AFMR phenomena. We anticipate that these new 
insights will facilitate improved interpretations and applications in spintronic and magnonics experiments.

\bibliography{manuscript}

\end{document}